\documentclass{sig-alternate-2013}

\usepackage{graphicx}
\usepackage{ifpdf}
\ifpdf
\fi
\usepackage{amsmath,amssymb}
\usepackage{subfigure}
\usepackage{url}
\usepackage{verbatim}
\usepackage{algorithm,algorithmicx}
\usepackage{algpseudocode}
\usepackage{times}
\usepackage{ifpdf}
 
\newcommand{\para}[1]{\smallskip \noindent {\bf #1}}
\newcommand\abs[1]{\left|#1\right|}

\newcommand{\latex}{\LaTeX}

\newcommand{\eat}[1]{}

\newfont{\mycrnotice}{ptmr8t at 7pt}
\newfont{\myconfname}{ptmri8t at 7pt}
\let\crnotice\mycrnotice%
\let\confname\myconfname%

\permission{Permission to make digital or hard copies of all or part of this work for personal or classroom use is granted without fee provided that copies are not made or distributed for profit or commercial advantage and that copies bear this notice and the full citation on the first page. Copyrights for components of this work owned by others than ACM must be honored. Abstracting with credit is permitted. To copy otherwise, or republish, to post on servers or to redistribute to lists, requires prior specific permission and/or a fee. Request permissions from permissions@acm.org.}
\conferenceinfo{CIKM'13,}{Oct. 27--Nov. 1, 2013, San Francisco, CA, USA.}
\copyrightetc{Copyright 2013 ACM \the\acmcopyr}
\crdata{978-1-4503-2263-8/13/10\ ...\$15.00.\\
http://dx.doi.org/10.1145/2508497.2508500 
}

\begin{document}

\title{First Author Advantage: Citation Labeling in Research
\titlenote{Authors by alphabetical order}}

\numberofauthors{3} 
\author{
\alignauthor
Graham Cormode\\
       \affaddr{University of Warwick}\\
       \email{G.Cormode@warwick.ac.uk}
\alignauthor
S. Muthukrishnan\\
        \affaddr{Rutgers University}\\
           \email{muthu@cs.rutgers.edu}
\alignauthor Jinyun Yan\\
       \affaddr{Rutgers University}\\
       \email{jinyuny@cs.rutgers.edu}
 }

\maketitle
\begin{abstract}
\small
Citations among research papers, and the networks they form, are the
primary object of study in scientometrics.  
The act of making a citation reflects the citer's knowledge of the
related literature, and of the work being cited. 
We aim to gain insight into this process by studying {\em citation
  keys}: user-chosen labels to identify a cited work. 
Our main observation is that the first listed author is
disproportionately represented in such labels, implying a strong mental bias
towards the first author. 
\end{abstract}


\section{Introduction}

The notion of a citation -- the formalized reference to a prior work
-- is at the heart of academic writing. 
No piece of work is complete without reference to related efforts, to
set the new contribution in context. 
Consequently, the study of citations is a central component of
understanding the relation between different articles. 
Indeed, the primary basis by which the impact of a piece of work is
assessed is by counting the number of citations that it has received. 
A large number of metrics for determining the importance of a
researcher, which rely on tracing citations in one way or another: 
$h$-index~\cite{hirsch2005index},
$g$-index~\cite{egghe2006improvement}, and many more.

There are a broad range of studies on citations. Besides using citations to measure the impact of a paper and the influence of an author, researchers build graphs of citations~\cite{gilbert1997simulation}, and study the structure, dynamics, and collaboration within and between academic fields. 
Leydesdorff and Amsterdamska~\cite{leydesdorff1990dimensions} analyzed the motivation behind a citation based on surveys to authors, and examined whether the cited and citing authors are in a professional relation, whether the citation behavior is for social or for cognitive purpose.

In this work, we take a look at the process of citation from a
different perspective.  
We focus on the process of an author making a citation, and ask what we
can learn about this act. 
To gain a vantage point on this process, we take advantage of the fact
that many researchers make use of computerized document preparation
systems. 
In particular, systems such as Bibtex and Endnote facilitate the
insertion of citations into documents. 
To refer to a particular work, the researcher must create a `key' for it. 
We identify these citation keys as objects of interest. 
We argue that the researchers' choice of citation key gives us an insight
into how they think about the work that they are citing.  
Moreover, we claim that key affects how they remember the work: if the
key includes one name out of several authors, we believe that this is
the name that the researcher most associates with the work. 

Due to increasingly collaborative research, there are many disputes
regarding the order of authors~\cite{amber2012}, because the order of
authors reflect the credit for contribution and authorship of the
paper. Teja \textit{et al.}~\cite{teja2007} showed several conventions
of author ordering, such as ranking authors by contribution levels and
arranging each group by alphabetical order; ranking authors strictly
by credit which declines with the position of authors; placing
important authors in the first and last position. 
Our work unveils the hidden citation keys, which to some extent
reflect how the researcher making a citation
thinks about the contribution of authors in a cited paper. 

To study the act of citation, we analyze a large data set of
\latex\ documents and their associated bibliographies. 
From these, we extract titles, years and names of authors in the
cited works, and measure how they relate to citation
keys. 
In total, we identify over 506K authors referenced in 225K citations
within 12K papers.  
We make a number of observations: 

\begin{itemize}
\item
\vspace*{-0.1in}
Most strikingly the first (listed) author of the paper is very
commonly included in the citation key.  
We argue that this gives a strong advantage to first authors, since
this creates a strong link between the first author and the paper. 
In particular, for areas which follow rules for author ordering
(e.g. alphabetical order), it can give strong benefits to authors who
are often listed first. 
\vspace*{-0.1in}
\item
Other authors are not neglected: we show that subsequent authors are
often referenced in the citation key, but less frequently and less
prominently.  
A common case is to list the first author by name, and subsequent authors by
initials. 
\vspace*{-0.1in}
\item We analyze the connection between citation keys and authors
  given context as author ordering, time and individual habit. We show
  that citing authors are less likely to favor first author if cited
  authors are in alphabetical order. 
We also observe a slightly declining ratio of using authors in
citation keys over time. 
Only a small portion of individuals stick to one labeling pattern when making many citations in a paper.
\vspace*{-0.1in}
\item
We study other frequently occurring terms in citation keys, and
observe that these include a variety of concepts: keywords indicative of the paper's content; descriptions of the type of the paper (article, thesis, book); and other meta-data such as the year of publication. 
\end{itemize}

Collectively, these give new insights into the nature of citation
keys, and perhaps into how researchers think about the works they
are citing. 

\section{Approach}
We focus on citation keys in \latex\ source files, 
which reveal the hidden citation keys. 
Our corpus consists of the 12,611 \latex\ sources for all papers
containing references from
computer science category  in `arXiv.org' up to April 2011.
More details about the dataset can be found in our prior work~\cite{CMY2012}.  Unfortunately, arXiv source files do not contain structured `bib' files but only have compiled unstructured `bbl' files.
References are either in a separate `.bbl' file or at the end of the
`.tex' file. We only consider papers that have references. 

Given unstructured bibliographies, we aim to extract citation keys, author names, title and year. It is a special case of the general Named Entity Recognition problem~\cite{nadeau2007survey}. However there is no existing labeled data to use for sophisticated learning algorithms. Therefore,
we adopt an approach which is easy-to-implement and achieves high precision and recall. 

We first introduce terminologies used through the paper.
The \emph{References} are a list of citations in a paper. 
A \emph{Bib Entry} is a citation in the references. 
A \emph{Citation Key} is the user-defined key to label the citation. 
\emph{Bib Meta} is a collection of meta information of a Bib Entry, such as, the citation key, author names, title, year, publishing organization, and so on. \emph{Author Text} is the piece of text in a \emph{Bib Entry} about authors. \emph{Authors} is a list of authors identified from \emph{Author Text}. \emph{Year} shows the publication year of the citation.

Our approach includes several steps. (1)  Bib entry identification: identify bib entries in references; (2)  Bib meta segmentation: segment the bib entry into pieces of meta info. (3) Author recognition: find author names (first and last names) in author text. 

\subsection{Bib Entry Identification}
In \latex , references are declared by \textbf{thebibliography} environment~\cite{wiki:biblio}. We first select the reference inside valid `thebibliography' environment, then extract bib entries in the reference. There are multiple commands to include a citation:  \textit{\textbackslash bibitem}; the \textit{\textbackslash bibitemstart} and  \textit{\textbackslash
  bibitemend} pair; and \textit{\textbackslash BIBentry}.  
We extract bib entries based on the usage of above commands. In total, we extracted \textbf{305,949} bib entries, on average each paper has \textbf{24.26} citations.

\begin{figure}[t]
\centering
\includegraphics[width=0.45\textwidth]{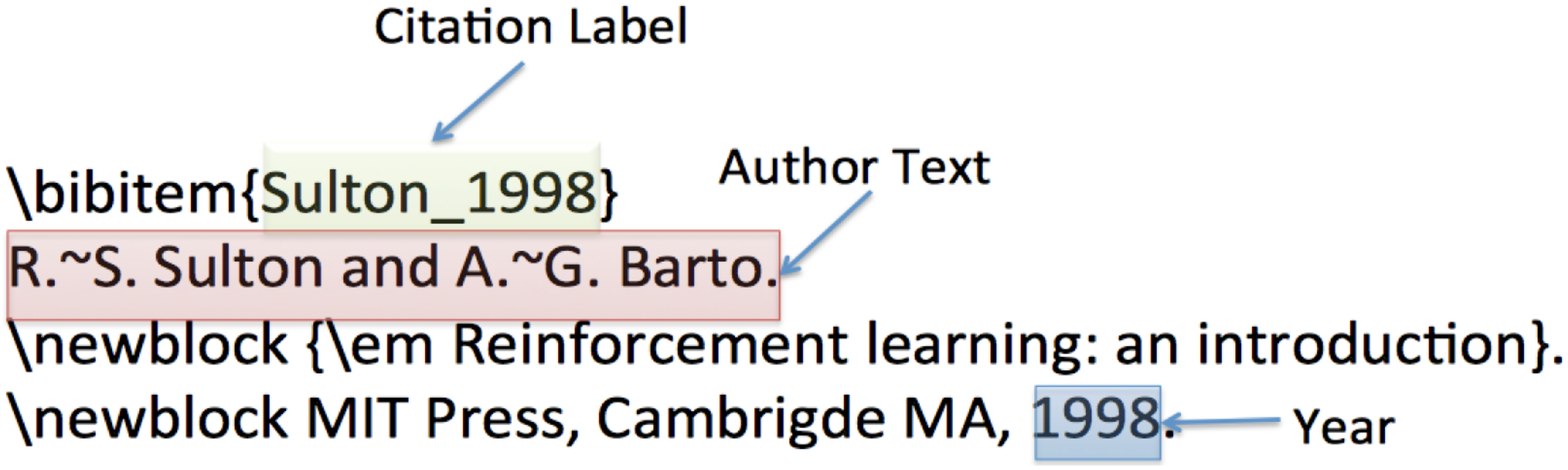}
\caption{Bib Meta Extraction}
\label{bibMeta}
\vspace*{-0.2in}
\end{figure}

\subsection{Bib Meta Segmentation}
For each bib entry, we want to extract meta data of the corresponding
citation. 
Here we focus on the citation key, authors, and year. 
Figure~\ref{bibMeta} shows an example of bib entry and its bib
meta. We describe steps to extract each piece of meta data in the following.

\para{Citation Key.}
Commands to include citations follow the format:
\centerline{\textbackslash $\langle$ command $\rangle$ [$\langle$ \textrm{explicit key} $\rangle$]\{ $\langle$\textrm{citation key} $\rangle$ \}}

The ``explicit  key'' is the index printed to identify the
citation. 
The number of explicit keys can be none or more than one; if omitted, the compiler automatically generates the citation index. The ``citation key'' is chosen, most of time, invented by authors. It is not be printed in the final paper, and is the unique key to match up the citation context in the paper and citation entry in the reference. 
We scanned the bib entry text to match the above patterns, and extracted \textbf{304,857} citation keys, which account for $99.6\%$ of bib entries. 

\para{Year.}
Some \latex\ commands can be used to identify publication year in the bib entry. For example, `bibinfo',`byear' and `bibyear'. 
If any of these commands exists, we use a regular expression to extract the year. However, only a small portion of bib entries use these
commands. To identify the publication year in bib entries, we search all four-digit terms which can be valid year candidates. 
The challenge is that these candidates can be volume number or page number, and cause false positive. For example, in Figure~\ref{bibMeta}, both ``1730'' and ``1999'' are candidates. 

\begin{figure}[t]
\centering
\includegraphics[width=0.45\textwidth]{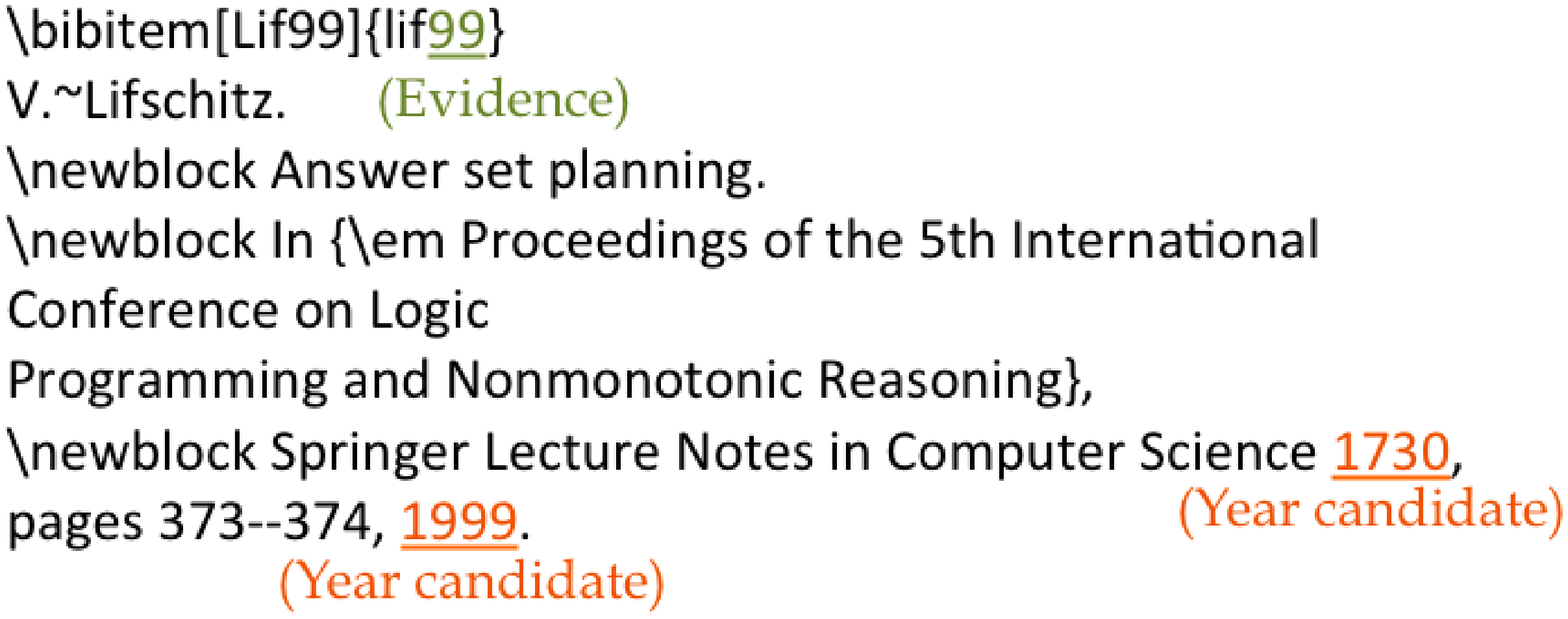}
\caption{Evidence based year extraction}
\label{bibMeta2}
\vspace*{-0.24in}
\end{figure}

An evidence-based algorithm is used to improve the accuracy of  year
extraction.  Observing that quite often citation keys contain digits about the year, we compare digits in citation keys and detected four-digit year candidates. As to the example in Figure~\ref{bibMeta2}, the citation key is
`lif99' which matches the candidate `1999'.  We focus on two-digit or
four-digit sub-strings in citation keys, and match them with year
candidates and return the matched candidate. 
There are quite a few cases where the publication year is one year
later or before the year marked in citation keys. 
We handle such cases by allowing a $\pm1$ variation of the matching.
If there is no evidence in the citation label, and there is more than
one year candidate, we will choose the first one.  By our
observation, year is usually put before publishing organization,
volume and page. 

\begin{table}[t]
\caption{Word features, examples and intuition}
\small
\centering
\begin{tabular}{| l |p{11mm}|p{38mm}|}
\hline
Word Feature & Example Words & Intuition\\
\hline
starts\_with\_brace & \{A\}spect& usually in title.   \\
\hline
ends\_with\_brace&  Theory\},& The end of title, when title is enclosed by braces. \\
\hline
has\_internal\_brace&  \{A\}spect & Part of the word has braces. \\
\hline
ends\_with\_comma& Fischer, & Punctuation: delimiter between last name and first name, or between semantic blocks, e.g., author text and title text \\
\hline
ends\_with\_period&    Y.~Singer. &Punctuation: the end of a semantic block, or initials.\\
\hline
capital\_period &  M. & Initials \\  
\hline
capital\_period\_dup& M.M., M.~M., M.-M. & Initials including middle names. \\ 
\hline
init\_capital&   Improved & Either last name or the beginning of the title. \\
\hline
four\_digit\_year& 2006&  Year \\
\hline
all\_alpha&  analysis &  Likely to be words in the middle of title.\\
\hline
all\_digits&    44 & Volume number, page number or other numbers.\\
\hline
all\_symbols & " & Braces, double quotes, single quote etc. \\
\hline
mixed\_case& ProSys, Cesa-Bianchi  & Self-defined system name, algorithm name, or author name. \\
\hline
all\_upper&  ACM & Special pronoun: self-defined system or algorithm name.  \\
\hline
all\_lower&  logic  & Usually word in the middle of title \\
\hline
internal\_symbol& Finite-time,  Cesa-Bianchi & Hyphen connects two words in title or in few author names.\\
\hline
token\_length& 4 & The number of characters in a word \\
\hline 
summarized\_pattern&  [1,~22021] & The pattern of the word\\
\hline
token\_word\_id & [1,~285901]  & Words removed symbols \\
\hline
\end{tabular}
\label{wordfeature}
\end{table}%

\para{Author Text.}
There is some prior work on identifying author text in a citation. 
Sarawagi {\em et al.}~\cite{sarawagi2003resolving} used hand-tuned
regular expression which exploit the pattern that single letter
initials before or after a word denoting last name. 
However, this method only handles one type of author names in
citation. The authors did not provide any results on the accuracy or
coverage of the method.  
We saw that in fact 
there are many different ordering of last name, first name and initials. 
Taking one broadly cited author for example, we found
many distinct variations on the ordering:
{\em ``D.~E. Knuth.''} ,  {\em ``Donald~E. Knuth''}, {\em
  ``D.~E.~Knuth''}, {\em ``Knuth, D.''}, {\em ``D. Knuth''}, {\em ``Knuth, D.~E.,''}, {\em ``D.~Knuth''}, {\em ``Knuth, D.E.''}, {\em ``D.E. Knuth''}.

When the author name contains two words and one initial, i.e., in the
form of ``Donald~E. Knuth'', it increases the difficulty to detect
author text without missing words in authors or including extra words
in title. 
The situation gets more complex if the bib entry has more than one
author, and each of them has a different ordering of first and last
names, as often occurs. 


We studied a large random sample of bib entries and chose $7$ patterns that can guarantee accurate extraction. These patterns accompany the \latex\ commands `newblock', `bibinfo', `Name', `bauthor',`bibsc', and so on.  
Patterns are matched in a fixed order, because some patterns have overlaps. 
Author text extracted by pattern matching are used as the ground truth
set.
We then design word features and adopt machine learning technique to
extract author text in the remaining citations. 
We consider word features that can distinguish author name and not-a-name words. Table~\ref{wordfeature} gives the detail of features we used. 

In Table~\ref{wordfeature}, the summarized pattern of a word is used
to cover the various usage of capital, non-capital and symbolic
characters. 
We summarize a word using the following heuristics. 
Internal symbols like hyphen are kept and unchanged.
\begin{center}
\begin{tabular}{rlp{1cm}rl}
 \textrm{[A-Z]} & $\rightarrow$ A & & \textrm{[a-z]} 	& $\rightarrow$ a  \\ 
 \textrm{[A\{a\}+]} & $\rightarrow$ Aa & & 
 \textrm{[\{0-9\}+]} & $\rightarrow$ d 
\end{tabular}
\end{center}

The token word is the word after removing symbols. The word length is
the length of the token word. We label each word in author text
extracted in ground truth set as instance in the ``NAME'' class (with
label 1). All words after the author text in a bib entry are instances in the ``NOT-A-NAME'' class (with label 0). We generated \textit{4,673,538} instances with features, which have \textit{942,789} ``NAME'', and \textit{3,699,852} ``NOT-A-NAME'' instances. Note that a small number of names are also in ``NOT-A-NAME'' class because some authors include editor names near title text.  

We apply logistic regression on word features we selected. We split
the labeled set into 7:3 train and test sets. We use 5-folder cross
validation and average results over 10 rounds. We obtain
\textbf{0.9276} precision, \textbf{0.9251} recall and \textbf{0.926}
F1-score. 
The result shows that features we selected are informative. 
We use the trained classifier to detect author text in unlabeled
citations. 
Our future plan to improve performance is to model the sequential
connection between words, using graphic models like conditional random
fields. 
Since we removed most \latex-specific commands from bib entries (some curly
brackets are kept), our approach can also be applied to bib entries
obtained from other sources e.g. from OCR scans of papers or from the web. 

\subsection{Authors Recognition}
The last step is to extract the list of author names from the detected author
text.  
Authors are not presented in a uniform format among entries, and
often vary even within the same entry. Some examples:

\begin{enumerate}
\item Example 1: \text{Partee,B.H., A. ter Meulen, and R.E. Wall}
\vspace*{-0.1in}
\item Example 2: \text{K.~Sagonas and T.~Swift and D.S. Warren.}
\vspace*{-0.1in}
\item Example 3: \text{Fillmore, C.J., P. Kay, and M.C. O'Connor.}
\end{enumerate} 

We use a combined heuristic and probabilistic method to separate author names in author text. 
We assume that a name can be partitioned into a first name and last
name. First name can be initials or a full first name. 
Middle names are always placed in between, thus we focus on detecting
the boundary between the first name and last name groups. 
The probabilistic method is used to identify whether a word is last name or first name, using word features. There are cases that our method failed to find the pair of first and last names. Most of them are names with single word for organization, institute, or software, e.g. ``Telelogic'',``Sun''. We set the first name to be empty and assign such words as last name. In total, we identified \textbf{506,634} authors from \textbf{225,438} entries.

\section{Findings}
\subsection{First Author Advantage}
We match the citation key and last names of authors by different
similarity metrics to find the connection between the authors' last names
and citation key.  
Let string $s_c$ be the citation key, and $s_a$ an author's last name.   
The function $f(s_a, s_c)$ returns $1$ if matched, and $0$ otherwise. 

\begin{table}[t]
\caption{The conditional probability of the $i$th author appearing in
  the citation key, given the number of authors}
\small
\centering
\begin{tabular}{|c|c|c|c|c|c|}
\hline
&Y = 1&2&3&4&5\\
\hline
1 &  0.62 &-&-&-&-\\
\hline
2 &  0.51 &0.17& -&- &-\\
\hline
3 & 0.48 & 0.09 &0.08&-&-\\
\hline
4 & 0.45 & 0.04& 0.05 &0.04&-\\
\hline
5 & 0.40 & 0.03 & 0.03& 0.04& 0.02\\
\hline
\end{tabular}
\label{condprob}
\end{table}%

\para{Exact matches.}
We consider exact substring matches first, i.e. a function $f$ so that 
if $s_a$ is a substring of $s_c$, $f(s_a,s_c) =1$.  
We find that the first author's last name has the most exact matches,
which covers $54.5\%$ of all citation keys.  
As the number of authors in papers ranges from one to many,
we study whether this high presence of the first author's name is
affected by the number of authors. 
A hypothesis to test is that with more authors, the ratio of matched first
author should decrease.  
We compute the conditional probability of citation keys matching
authors in each position, given the number of authors. 
We use random variable $X= \{1,2,3, \ldots\}$ to represent the number
of authors, and $Y = \{1,2,3,\ldots\}$ represent the matched author
position. 
The conditional probability is computed as 
\[\Pr[Y= i | X=j ] = \frac{\Pr[Y=i, X=j]}{\Pr[X=j]},\] 
where $\Pr[X=j]$ is the probability of $j$ authors in the
citation paper, and $\Pr[Y=i, X=j]$ is the probability that citation key matches the $i-$th author in a paper with $j$ authors.
Table~\ref{condprob} shows the result for $X= 1,2,3,4,5$. The row
shows the value of  the number of authors $X$,  and the column in the
table shows the matched author position $Y$. We can see that as the
number of authors increases, the ratio of matching first author last
name decreases only slightly. 
The chance that authors in higher positions are mentioned in citation keys does not increase with the number of authors.  

The result shows that the first author is
dramatically more likely to be included in the citation key than any
subsequent author. 
For a two author paper, it is three times more likely that the first
author will be identified in the key. 
For four or five author papers, the first author is approximately ten
times more likely to be identified than any one of the subsequent
authors.  
Although the first author's presence decreases with the number of
authors, it remains high (40\%). 
This supports the notion of ``first author advantage'': the
idea that the first named author is much more strongly associated with
the work than subsequent authors. 
This seems to hold, at least as far as presence in citation key is
concerned.  
We study this issue further in our subsequent experiments. 

\begin{table}[t]
\caption{Similarity Metrics $f(s_a,s_c)$}
\centering
\small
\begin{tabular}{|l|l|c|}
\hline
Id &Metric Name & Description  \\
\hline
M1&Exact Match &   $\operatorname{substring}(s_a, s_c)$  \\ 
\hline
M2& Longest Sequence Ratio &  \large{$\frac{\operatorname{lcs}(s_a,s_c)}{|s_a|}$ }   \\ 
\hline
M3&$n$-Gram Jaccard Similarity &  \large{$\frac{|S_a \cap S_c|}{|S_a \cup S_c|} $}\\ 
\hline
M4&$n$-Gram asymmetric Similarity &  \large{$\frac{(S_a \cup S_c)^w - (S_a \setminus S_c)^{w}}{(S_a \cup S_c)^w}$}\\
\hline
M5&$n$-Gram Dice Coefficient & \large{ $\frac{2 |S_a \cap S_c| }{|S_a|+|S_c|}  $}\\
\hline
\end{tabular}
\vspace*{-0.1in}
\label{similar:metric}
\end{table}%

\para{Approximate matches.} 
In some cases, citation keys contain fragments of last name of
authors, rather than full names. 
To study this further, 
in addition to exact match, we use several other similarity metrics to
estimate whether the citation key matches with an author's last name. 
Metrics we used are described in Table~\ref{similar:metric}. 
In the table, $S_a$ is the set of $n$-grams of the author string $s_a$, and
$S_c$ for citation key string $s_c$.
We make use of the length of the longest common substring ($\operatorname{lcs}$)
between two strings, and use $|\cdot|$ notation to denote the length
of a string or the size of a (multi)set. 

For $n$-Gram based metrics, we set $n$ to be $\min(3,|s_c|, |s_a|)$. 
We set the threshold between matched and un-matched to be $0.5$ for
all metrics. 
For the weighted $n$-Gram asymmetric similarity metric, the weight
parameter $w$ is learned from manually labeled samples, as  $0.5$.
Each metric has some advantages and disadvantages. 
For example, Jaccard coefficient will lead to a false negative if the citation key is much longer than
last name, i.e. when the citation key contains both author's last name
and title words. 

To have a clear understanding of the performance of these metrics, we
manually analyzed and labeled a random selection of 432 instances. 
If the human judge determined the key is based on an author's
last name, the example is labeled as positive, even for the case that only the
first letter of last name is used. 
For example, for a paper with two
authors, with last names ``Ladner'' and ``Reif'' respectively, if the
citation key is ``LR'', we label the key as matching both authors' last
names. 

\begin{table}[t]
\caption{The Performance of Similarity Metrics}
\centering
\small
\begin{tabular}{|c|c|c|c|}
\hline
Metric ID & precision & recall&F1\\
\hline
M1 &1  & 0.48 & 0.65\\
\hline
M2 & 0.99& 0.41& 0.58\\
\hline
M3 & 1&0.19 &0.32\\
\hline
M4  & 0.77 &0.76 &0.76\\
\hline
M5 & 1 & 0.32&0.48\\
\hline
\end{tabular}
\vspace*{-0.1in}
\label{sim:performance}
\end{table}%

Table~\ref{sim:performance} shows the performance of each metric
on the labeled data. The result shows that exact match performs well but $n$-Gram asymmetric similarity has the best
performance. 
We thus use the $n$-Gram asymmetric similarity metric to estimate the matching between
citation keys and authors in each position, and it shows
\textbf{61.8\%} citation keys matched the first author, \textbf{21\%} for
the second author, \textbf{10\%} for the third author, \textbf{4\%}
and \textbf{1.5\%} for the fourth and fifth author
respectively. 
Thus, even allowing approximate matches and abbreviations, there is
still strong evidence for a first author advantage -- all positions
increase their likelihood of matching compared to seeking exact
matches, but the first author is still much more likely to be
referenced in the citation key. 


\subsection{Author Acronyms}
The above metrics measure whether the last name of the author in each
position is used in the citation labels. 
However, there are cases which these will miss, such as 
where the key uses the acronym of last names
of all authors as the citation key, e.g. ``CMY''; or uses the last name of
the first author and initials of the rest, e.g. ``CormodeMY''; or
the first few characters of each author's last name, e.g. ``CorMutYan''. 
Digits and other title words might also be attached to these patterns. 
To detect the presence of such acronym patterns,  we
used a weighted longest common sub-sequence algorithm. 
Algorithm~\ref{wlcs} gives the details of the matching. 

\begin{algorithm}[t]
\caption{Matching Author Acronym }
\label{wlcs}
\begin{algorithmic}
\State \text{Input: list of last names $L$, citation label $s_c$,}
\State \text{ 	string of all last names $ s_A $ }
\State \text{Output: True if $s_c$ is the acronym of $s_A $, false otherwise}
\Function{IsAcronym}{($s_c, s_A, L, w_A, w_a, w_s $)}
	\State $ T = len(L)$. 
	\State  Let $A$ be $(\abs{s_c}+1)\times(\abs{s_A}+1)$ matrix of zeros.
	\For { $i = 1 \to  \abs{s_c}$  }
		\For{ $j = 1  \to \abs{s_A }$} 
			\State $p \leftarrow \max(A[i-1,j], A[i,j-1])$
			\State $w = 0$
			\If{ $\operatorname{lower}(s_c[i-1]) == \operatorname{lower}(s_A[j-1]) $} 
				\[
				w = 
				\begin{cases}
				w_A, &\text{$s_A[j-1]$ is Capital First letter }\\
				w_a, &\text{$s_A[j-1]$ is Non Capital, First letter }\\
				w_s, &\text{otherwise}
				\end{cases} 
				\]
			\EndIf
			\State $A[i,j] = \max(p, A[i-1,j-1]+ w)$
		\EndFor
	\EndFor
	\State $s \leftarrow A[\abs{s_c},\abs{s_A}]$  \Comment{Score for the best match}
	\State \Return $(s > 0.5 * w_A * T)$
\EndFunction
\end{algorithmic}
\end{algorithm}

We define $s_c$ and $s_A$ be the citation key and the string concatenating the last names of all authors, respectively. 
We modify the longest common sub-sequence algorithm by assigning
weights to letters in different position of the author string. 
The final score is then used to decide if the citation label is the
acronym of all authors. 
We assign scores in the following way. 
If the matched letter is the first letter of an author's last name,
and the letter is a capital letter, 
we assign $w_A = 2 $;  if it is not a capital letter,  $w_a = 1.1$. 
We assign $w_s = 0.1$ to other matched letters, i.e. non-initial
letters of last names. 
The score $s$ of a match is the sum of weighted score of matched
letters. We test the score against the threshold $s > 0.5*w_A*T$ where $T$ is
the number of authors and $0.5$ is the minimum threshold if lowercase
acronym is matched. If the score is larger than the threshold, the
matching function returns \emph{True}, otherwise it returns \emph{False}. 

\begin{figure}[t]
\centering
\includegraphics[width=0.45\textwidth]{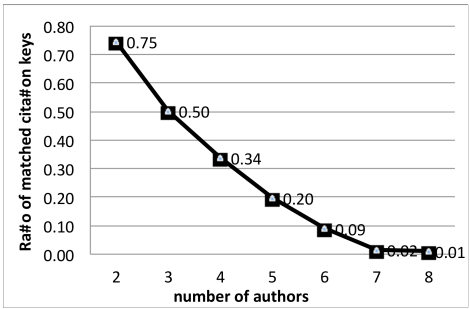}
\caption{Ratio of author acronyms, given the number of authors}
\label{authoracronym}
\vspace*{-0.1in}
\end{figure}

By this algorithm, we found \textbf{72\%} 
of the citation keys contain an acronym of last names of all
authors. 
Note that our method will typically match the case that one or more
full last names are in the citation key. 
When we exclude the single-author cited papers 
the ratio of author acronym usage is \textbf{56\%}. 
The matching algorithm reported that at least half of citation keys
covered all authors, even though usually the full last name of authors
in later positions is not included.  
This still leaves almost half of citation labels that have some
different meaning and format, which are interesting to explore.  

Figure~\ref{authoracronym} shows 
the ratio of citation keys which are an acronym of all authors 
as the number of authors increases.
We observe a strictly decreasing line which shows that with more
authors, the likelihood of using acronym of all authors reduces. 
However, comparing to the conditional probability of exact matching an author in high position, the coverage of all authors is much higher.   
From this we conclude that as there are more authors on a paper, the
chance of each author getting referenced in the citation label (and
so, we conjecture, figuring highly in the thoughts of the researcher
making the citation) becomes lower. 
Moreover, the chance of being referenced falls with position: we still
observe a much greater chance for the first ordered author to be referenced than
any subsequent author, even when we consider acronyms. 

\subsection{Labeling Behavior}
We analyzed the connection between the position of authors and
citation keys. 
Here we include some context of labeling behavior, in
particular, the order of authors in cited papers and the time. 
We also examine whether creators of citations follow a fixed pattern. 

\para{Author Ordering.} We analyze whether the order of authors in
citations affects citation keys. 
When authors are not listed in alphabetic order, 
it is common to rank authors by their contribution, so the
first author has strongest ownership of the paper. 
When authors are placed in alphabetic order, chances are that the main
contributor is not in the first position. 
Table~\ref{order} shows the breakdown of cases based on the ordering
of authors (alphabetic or non-alphabetic), then by
(1) whether the citation key references all authors;
and 
(2) whether the first author's name is given in full. 
Note that the ratio is computed by number of citation keys that
matched the criteria to the total number of citation keys, and 
only citations with more than one author are considered.  
For (1), we see that when authors are listed alphabetically, the
citation key are twice as likely to reference all authors. 
But when authors are not listed alphabetically, there is no great
difference between the cases. 
For (2), when the authors are in alphabetical author, the key is more
likely to not invoke the first author in full.
But when not in alphabetical order, the key is more likely to invoke
the first author. 

The results in Table~\ref{order} supports the idea that 
 when alphabetic order is used, 
people are aware that the first author is not necessarily the main
contributor and thus are more likely to touch on all authors, and less
likely to explicitly mention them.

\begin{table}[t]
\caption{ Author ordering and citation keys, at least 2 authors }
\centering
\small
\begin{tabular}{|p{3.3cm}|p{2cm}|p{2cm}|}
\hline
&author acronym match = True  &author acronym match = False\\
\hline
in alphabetical order  & 43\% & 19.6\%\\
\hline
not in alphabetical order  & 18\%&19.5\%\\
\hline
\hline
&first author match = True  &first author match = False\\
\hline
in alphabetical order  & 29.5\% & 33.2\%\\
\hline
not in alphabetical order  & 21\%&16\%\\
\hline
\end{tabular}
\label{order}
\vspace*{-0.1in}
\end{table}%
 
\para{Trend over Time.}
We conduct preliminary analysis on whether the labeling pattern
changes over time.  We select papers published across two decades: 
(1990, 2000) and (2000, 2010), and compute the ratio of citation keys
with author acronym pattern and first author exact matching
pattern. Table~\ref{trend} shows the trend.  
More citations are made between (2000, 2010), in part because of the
increasing number of papers published over time~\cite{larsen2010rate}. 
Interestingly, we observe an appreciable declining trend for both
labeling patterns. 
This implies that over time, people are less commonly using author
names in citation keys. 
Our next step will be to gather more data across time and examine the trend over longer time periods, and explore other patterns of citation keys.

\begin{table}[t]
\caption{Labeling pattern over time}
\centering
\small
\begin{tabular}{|l|l|l|}
\hline
&(1990, 2000) & (2000, 2010)\\
\hline
number of citations & 75568 & 123188\\
\hline
number of author acronym matches & 54\% & 46.1\%\\
\hline
number of first author matches  & 40.8\% &37.5\% \\
\hline
\end{tabular}
\vspace*{-0.1in}
\label{trend}
\end{table}%

\para{Consistency.}
We next consider the consistency of formation of citation labels:
when an author writes a paper, will he follow the same pattern of citation keys for all citations? 
We examine two patterns: exact matching the first author's last name, and
approximately matching an acronym of authors. 
We compute the pattern matching ratio of the paper $p_i$ by 
\[ \operatorname{pmr}(p_i) =  \frac{f\text{(citations with keys matching pattern)}}{f\text{(citations)} }\]
If $\operatorname{pmr}(p_i) =1$, the authors consistently follow one
pattern across all citations in the paper.  
We found \textbf{20.4\%} of papers consistently use author acronym pattern, and \textbf{12.6\%} of papers follow exact matching first author's last name pattern. If we set the consistency ratio to $0.9$, there are \textbf{32\%} and  \textbf{20.8\%} of papers ``mostly consistently'' using author acronym pattern and first author last name respectively.
These ratios are low,  indicating that authors use various methods to compose citation keys. 
A conjecture is that when papers are written by multiple authors,
variations in citation keys are introduced by coauthors' different
habits, and the lack of incentive to make them consistent.

\subsection{$n$-Gram Analysis}
The above results show the last name(s) of author(s) 
are present in some form or another in a majority of 
citation keys. We still have a large portion of citation keys that are not
related to author names. To understand the meaning hidden in a citation key, we leverage $n$-Gram analysis. Contiguous sequences of $n$ characters from $304,857$  citation keys are computed. Here $n$ ranges from $2$ to $10$, and all citation keys are lowercased. 

\begin{figure}[t]
\centering
\includegraphics[width=0.36\textwidth]{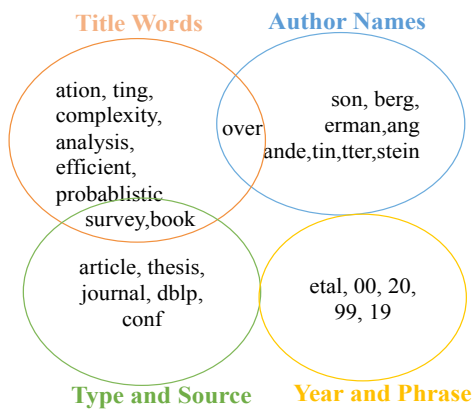}
\caption{$n$-Grams Meaning Clusters}
\vspace*{-0.15in}
\label{meaningcluster}
\end{figure}

We select top 20 frequent $n$-grams with $n$ from $2$ to $10$. We first observe that digits representing years are very frequent in
citation keys, as well as the phrase ``et al'' to mean ``and other authors''.   We cluster them into four pre-defined meaning clusters: ``title words'', ``author names'', ``type and sources'', and ``year and phrase''.  Figure~\ref{meaningcluster} illustrates several hand-picked terms in each meaning group. Terms in ``type and sources'' and ``year and phrase'' clusters are measured by human judges. Terms in ``author names'' and ``title words'' are measured by the affinity of a term to author names versus to titles. The affinity to title is the ratio of the number of times the term in title to the total occurrence of the term. The affinity to authors is computed similarly. 

Some terms have close affinity scores to two
clusters. An example is the term ``over'', which is present in both author names, i.e. ``Cover1991'',  and title words, i.e. ``zhu:coverage''.  Similarly, the term ``survey'' is common in both ``type and resource''
and ``title words'', since many survey papers use titles as ``A survey
of ...''. Except these pre-defined four clusters, there are other meaning groups of citation keys, e.g., conference abbreviations. However, the amount of citation keys containing a particular conference abbreviation is not large enough. To understand the behavior of including conference name in citation keys, we can match citation keys with a list of all conference abrreviations. Such further analysis of the semantics of citation keys is left to the future work. 
  



\para{Use of Bibliographic Resources.}
There are many available sources of bibliographic information. 
For example, many journal websites allow the user to export a reference
in Bibtex format, for use in their own papers. 
We found an interesting copying behavior due to the frequent occurrence of the term ``dblp''. Some authors directly copy the bib entry from DBLP, which labels citations with a fixed format. 
Around 0.5\% of all citation keys in our dataset are copies from DBLP: a fraction small enough that it does not alter any of our above
analysis, but still appreciable. 
For papers with DBLP copies, around 4\% of them contain more than half
bib entries copied from DBLP.  We also find such copying behavior occurs more common in recent decade. \textbf{70\%} of  DBLP citation keys belong to papers later than the year 2000. For papers having DBLP copied citation keys, \textbf{86\%} of them are later than the year 2000.  

It is possible that more researchers copy from DBLP, but 
modify the citation keys to fit their own habits, which reduces the value of the observed ratio. However, it is infesible to identify such cases by only taking advantage of bbl files in our dataset. 
For more recent papers, there are entries examples taken from 
Google scholar, which labels papers by last name of first author, year and first word in the title. If we revisit this problem 10 years later, it will be interesting to examine whether the labeling pattern converges to copying from such bibliography services. 
\vspace{-0.7em}
\section{Concluding Remarks}
 
The use of citation keys offers a rare insight into the process of
making a citation, and gives some perspective into how the researcher
making the citation thinks of the work being referenced. 
We have seen that there is a dramatically strong occurrence of the
first (listed) author in such keys, far more so than other authors. 
We conjecture that this indicates that the first author receives much
greater prominence than other authors.
This ``first author advantage'' may have many consequences, particularly in disciplines
which follow a rule (such as alphabetical ordering of authors) that
mean certain researchers have a much higher chance of being listed
first. 

There are many further questions to address around questions of citation
and citation labeling.  We've analyzed how citation keys relate to
authors in each position. It will be interesting to investigate
reasons why people sometimes pick authors that are not listed first.  
We have indicated some ways in which citation labels are formed (from
authors, from topics, from venues and from years), but it remains to
fully understand these different sources, and to study what impact
these have on how the work is thought of. 
A direction for further work is to study the citation label in the
context of the citation: if we analyze the text of the paper where the
citation is made, can we determine if the sentiment towards the cited
work is positive or negative?  
Does the label give further insight into the researcher's feelings
towards the cited work's importance or depth?
%

\bibliographystyle{abbrv} 
{\small
\bibliography{scienceography}

\begin{thebibliography}{10}

\bibitem{wiki:biblio}
\url{ http://en.wikibooks.org/wiki/LaTeX/Bibliography_Management}.

\bibitem{CMY2012}
G.~Cormode, S.~Muthukrishnan, and J.~Yan.
\newblock {\em Scienceography: the study of how science is written}.
\newblock the Sixth International conference on Fun with Algorithms, 2012.

\bibitem{amber2012}
A.~Dance.
\newblock {\em Authorship: Who's on first?}
\newblock Nature, 2012.

\bibitem{egghe2006improvement}
L.~Egghe.
\newblock An improvement of the h-index: The g-index.
\newblock {\em ISSI newsletter}, 2(1):8--9, 2006.

\bibitem{gilbert1997simulation}
N.~Gilbert.
\newblock A simulation of the structure of academic science.
\newblock {\em Sociological Research Online}, 1997.

\bibitem{hirsch2005index}
J.~E. Hirsch.
\newblock {\em An index to quantify an individual's scientific research
  output}, volume 102.
\newblock National Academy of Sciences, 2005.

\bibitem{larsen2010rate}
P.~O. Larsen and M.~von Ins.
\newblock The rate of growth in scientific publication and the decline in
  coverage provided by science citation index.
\newblock {\em Scientometrics}, 84:575--603, 2010.

\bibitem{leydesdorff1990dimensions}
L.~Leydesdorff and O.~Amsterdamska.
\newblock Dimensions of citation analysis.
\newblock volume~15, pages 305--335. Sage Publications, 1990.

\bibitem{nadeau2007survey}
D.~Nadeau and S.~Sekine.
\newblock A survey of named entity recognition and classification.
\newblock {\em Lingvisticae Investigationes}, 30(1):3--26, 2007.

\bibitem{sarawagi2003resolving}
S.~Sarawagi, V.~V. Vydiswaran, S.~Srinivasan, and K.~Bhudhia.
\newblock Resolving citations in a paper repository.
\newblock {\em ACM SIGKDD Explorations Newsletter}, 5(2):156--157, 2003.

\bibitem{teja2007}
T.~Tscharntke, M.~E. Hochberg, T.~A. Rand, V.~H. Resh, and J.~Krauss.
\newblock {\em Author Sequence and Credit for Contributions in Multiauthored
  Publications}.
\newblock PLoS Biol, 2007.

\end{thebibliography}
}
 
\end{document}